\documentclass[11pt]{article} 
\usepackage[normalem]{ulem} 
\usepackage{amssymb} 
\usepackage{amsmath} 
\usepackage{amsthm} 
\usepackage{setspace} 
\usepackage{cite} 
\theoremstyle{plain}
\usepackage[hmargin=2.4cm,vmargin=2.9cm]{geometry} 
\newcommand\numberthis{\addtocounter{equation}{1}\tag{\theequation}} 
\newcommand{\NW}{\mbox{\small\rm NW}} 
\newcommand{\IMM}{\mbox{\small\rm IMM}} 
\newcommand{\VP}{\mbox{\small\rm VP}} 
\newcommand{\VNP}{\mbox{\small\rm VNP}}

\newcommand{\LM}{\mbox{\small\rm LM}} 
\newcommand{\ex}{\mbox{\small\rm X}} 
\newcommand{\ey}{\mbox{\small\rm Y}} 

\newcommand{\GL}{\mbox{\small\rm GL}}
\newcommand{\el}{L}

\newcommand{\F}{\mathbb{F}} 
\newcommand{\fq}{\mathbb{F}_q} 
\newcommand{\Q}{\mathbb{Q}}
\newcommand{\M}{\mathcal{M}}

\newcommand{\depththree}{\Sigma\Pi\Sigma} 
\newcommand{\depthfour}{\Sigma\Pi\Sigma\Pi} 
\newcommand{\depthfoursqrtn}{\Sigma\Pi^{[O(\sqrt{n})]}\Sigma\Pi^{[\sqrt{n}]}} 
\newcommand{\depthfourparam}{\Sigma\Pi^{[D]}\Sigma\Pi^{[t]}} 

\newtheorem{theorem}{Theorem} 
\newtheorem{definition}[theorem]{Definition} 
\newtheorem{question}[theorem]{Question}
\newtheorem{corollary}[theorem]{Corollary} 
\newtheorem{prop}[theorem]{Proposition} 
 
\newtheorem{lemma}[theorem]{Lemma}

\DeclareMathOperator{\var}{\mbox{\small\rm var}} 
\DeclareMathOperator{\codim}{\mbox{\small\rm codim}}

\DeclareMathOperator{\poly}{\mbox{\small\rm poly}}

\title{On the Limits of Depth Reduction at Depth 3 Over Small Finite Fields} 
\author{Suryajith Chillara \qquad Partha Mukhopadhyay \\ \small{Chennai Mathematical Institute, India}\\\small{\{suryajith, partham\}@cmi.ac.in}} 
 
\begin{document} 
\maketitle 
 
\abstract{
In a surprising recent result, Gupta et.al. \cite{gkks2013}  have proved that over $\Q$ any $n^{O(1)}$-variate and $n$-degree polynomial in $\VP$ can also be computed by  a depth three $\depththree$ circuit of size $2^{O(\sqrt{n} \log^{3/2}n)}$.  Over fixed-size finite fields, Grigoriev and Karpinski proved that any $\depththree$ circuit that computes the determinant (or the permanent) polynomial of a $n\times n$ matrix must be of size $2^{\Omega(n)}$. In this paper, for an explicit  polynomial in $\VP$ (over fixed-size finite fields), we prove that any $\depththree$ circuit  computing it must be of size $2^{\Omega(n\log n)}$. The explicit polynomial that we consider is the iterated matrix multiplication polynomial of $n$ generic matrices of size $n\times n$.  The importance of this result is that over fixed-size fields there is \emph{no depth reduction technique} that can be used to compute all the $n^{O(1)}$-variate and $n$-degree polynomials in $\VP$  by depth 3 circuits of size 
$2^{o(n\log n)}$. 
The result of \cite{gk1998} can only rule out such a possibility for $\depththree$ circuits of size $2^{o(n)}$. 

We also give an example of an explicit polynomial ($\NW_{n,\epsilon}(\ex)$) in $\VNP$ (which is not known to be in $\VP$), for which any $\depththree$ circuit computing it (over fixed-size fields) must be of  size $2^{\Omega(n\log n)}$. The polynomial we consider is constructed from the combinatorial design of Nisan and Wigderson \cite{nw1994} and is closely related to the polynomial considered in \cite{kss2013}. 
An interesting feature of our depth 3 lower bound results is that we provide the first examples of two polynomials (one in $\VP$ and one in $\VNP$) such that they have provably stronger circuit size lower bounds than Permanent in a reasonably strong model of computation, i.e. $\depththree$ circuits over fixed-size finite fields.  

Next, we explore the depth 4 circuit complexity of the polynomial $\NW_{n,\epsilon}(\ex)$ and prove that (over any field) any 
depth four $\depthfoursqrtn$ circuit computing it must be of size $2^{\Omega(\sqrt{n}\log n)}$.   
Before our result, Kayal et.al. \cite{kss2013} showed a depth four $2^{\Omega(\sqrt{n}\log n)}$ circuit size lower bound for an explicit polynomial in $\VNP$ and Fournier et.al. \cite{flms2013} showed a similar circuit size lower bound for a polynomial in $\VP$ (which is again the iterated matrix multiplication polynomial). 
The polynomials considered in \cite{kss2013} and \cite{flms2013} have a matching depth four $\depthfoursqrtn$ circuit size upper bound of $2^{O(\sqrt{n}\log n)}$. To the best of our knowledge, the polynomial $\NW_{n,\epsilon}(\ex)$ is the first example of an explicit polynomial in $\VNP$ such that it requires $2^{\Omega(\sqrt{n}\log n)}$ size depth four $\depthfoursqrtn$ circuits, but no known matching upper bound.

} 

\section{Introduction} 
\label{Section:Intro} 
In a recent breakthrough, Gupta et.al. \cite{gkks2013} have proved that over $\Q$, if an $n^{O(1)}$-variate polynomial of degree $d$ is computable by an  arithmetic circuit of size $s$, then it can also be computed by a depth three $\Sigma\Pi\Sigma$ circuit of size $2^{O(\sqrt{d \log d \log n \log s})}$. As a corollary of this result, they get a $\Sigma\Pi\Sigma$ circuit of size $2^{O(\sqrt{n} \log n)}$ computing the determinant polynomial of a $n\times n$ matrix (over $\Q$). Before this result, no depth 3 circuit for Determinant of size smaller than $2^{O(n\log n)}$ was known (over any field of characteristic $\neq$ 2).  
 
The situation is very different over \emph{fixed-size finite fields}. Grigoriev and Karpinski proved that over fixed-size finite fields, any  depth 3 circuit for the determinant polynomial of a $n\times n$ matrix must be of size $2^{\Omega(n)}$ \cite{gk1998}. Although Grigoriev and Karpinski proved the lower bound result only for the determinant polynomial, it is a folklore result that some modification of their argument can show a similar depth 3 circuit size lower bound for the permanent polynomial as well \footnote{Saptharishi gives a nice exposition of this result in his unpublished survey and he attributes it to Koutis and Srinivasan \cite{rp:pc}.}. Over any field, Ryser's formula  for Permanent gives a $\Sigma\Pi\Sigma$ circuit of size $2^{O(n)}$ (for an exposition of this result, see \cite{feige2009}).  Thus, for the permanent polynomial the depth 3 complexity (over fixed-size finite fields) is essentially $2^{\Theta(n)}$.
 
The result of \cite{gkks2013} is obtained through an ingenious depth reduction technique but their technique is tailored to the fields of zero characteristic. In particular, the main technical ingredients of their proof are the well-known monomial formula of Fischer \cite{fis1994}  and the duality trick of Saxena \cite{sax2008}. These techniques do not work over finite fields. Looking at the contrasting situation over $\mathbb{Q}$ and the fixed-size finite fields, a natural question is to ask whether one can find a new depth reduction technique over fixed-size finite fields such that any $n^{O(1)}$-variate and degree $n$ polynomial in $\VP$ can also be computed by a $\depththree$ circuit of size $2^{o(n\log n)}$. 

\begin{question}
Over any fixed-size finite field $\F_q$ for $q\geq 3$, is it possible to compute any $n^{O(1)}$-variate 
and $n$-degree polynomial in $\VP$ by a $\depththree$ circuit of size $2^{o(n\ln n)}$ ? 
\end{question}

Note that any $n^{O(1)}$-variate and $n$-degree polynomial can be trivially computed by a $\depththree$ circuit of size $2^{O(n\log n)}$ by writing it explicitly as a sum of all $n^{O(n)}$ possible monomials.  

We give a negative answer to the aforementioned question by showing that over fixed-size finite fields, any $\depththree$ circuit computing the iterated matrix multiplication polynomial (which is in $\VP$ for any field) must be of size $2^{\Omega(n\log n)}$ (See Subsection~\ref{Subsection:IMM}, for the definition of the polynomial).  More precisely, we prove that any $\depththree$ circuit computing the iterated matrix multiplication polynomial of $n$ generic $n\times n$ matrices (denoted by 
$\IMM_{n,n}(\ex)$), must be of size $2^{\Omega(n\log n)}$.  

Previously, Nisan and Wigderson \cite{nw1997} proved a size lower bound of $\Omega(n^{d-1}/d!)$ for any homogenous $\depththree$ circuit computing the iterated matrix multiplication polynomial over $d$ generic $n\times n$ matrices. Kumar et.al. \cite{kms2013} improved the bound to $\Omega(n^{d-1}/2^d)$. These results work over any field. Over fields of zero characteristic, Shpilka and Wigderson proved a near quadratic lower bound for the size of depth 3 circuits computing the trace of the iterated matrix multiplication polynomial \cite{sw2001}.

Recently Tavenas \cite{tav2013}, by improving upon the previous works of Agrawal and Vinay\cite{av2008}, and Koiran \cite{koi2012} proved that any $n^{O(1)}$-variate, $n$-degree polynomial in $\VP$ has a depth four $\depthfoursqrtn$ circuit of size $2^{O(\sqrt{n}\log n)}$. Subsequently, Kayal et.al. \cite{kss2013} proved a size lower bound of $2^{\Omega(\sqrt{n}\log n)}$ for a polynomial in $\VNP$ which is constructed from the combinatorial design of Nisan and Wigderson \cite{nw1994}. In a beautiful follow up result, Fournier et.al. \cite{flms2013} proved that a similar lower bound of $2^{\Omega(\sqrt{n}\log n)}$ is also attainable by the iterated matrix multiplication polynomial (see \cite{cm2013}, for a unified analysis of the depth 4 lower bounds of \cite{kss2013} and \cite{flms2013}). The main technique used is \emph{the method of shifted partial derivatives} which was used to prove $2^{\Omega(\sqrt{n})}$ size lower bound for $\depthfoursqrtn$ circuits computing Determinant or Permanent polynomial \cite{gkks2012}. Recent work of Kumar and Saraf \cite{ks2013} shows that the depth reduction as shown by Tavenas \cite{tav2013} is optimal even for the homogenous formulas. This strengthens the result of \cite{flms2013} who proved the optimality of depth reduction for the circuits. 

Similar to the situation at depth 4, we also give an example of an explicit $n^2$-variate and $n$-degree polynomial in $\VNP$ (which is not known to be in $\VP$) such that over fixed-size finite fields, any depth three $\Sigma\Pi\Sigma$ circuit computing it must be of size $2^{\Omega(n \log n)}$.  This polynomial family, denoted by  $\NW_{n,\epsilon}(\ex)$ (see Subsection \ref{Subsection:NW}, for the definition of the polynomial) is closely related to the polynomial family (with a small modification) introduced by Kayal et.al. \cite{kss2013}.  In fact, from our proof idea it will be clear that the strong depth 3 size lower bound results that we show for $\NW_{n,\epsilon}(\ex)$ and $\IMM_{n,n}(\ex)$ polynomials, are not really influenced by the fact that the polynomials are either in $\VNP$ or $\VP$. Rather, the bounds are determined by a combinatorial property of the subspaces generated by a set of carefully chosen derivatives. One interesting conclusion (which is somewhat counter intuitive) of the depth 3 circuit size lower bound results is that, we get the first examples of explicit polynomials (one in $\VP$ and one in $\VNP$) such that they have provably stronger lower bounds than Permanent in a reasonably strong model of computation, i.e. $\depththree$ circuits over fixed-size finite fields. 
 Our main theorem is the following. 
\begin{theorem}\label{main-thm}
Over any fixed-size finite field $\fq$ such that $q\geq 3$, any depth three $\depththree$ circuit computing the polynomials $\NW_{n,\epsilon}(\ex)$ or $\IMM_{n,n}(\ex)$ must be of size at least $2^{\delta n\log n}$,  where $\epsilon, \delta\in (0,1)$ and depend only on $q$.  
\end{theorem}

As an important consequence of the above theorem, we have the following corollary.
\begin{corollary}
Over any fixed-size finite field $\F_q$ for $q\geq 3$, there is no depth reduction technique that can be used to compute all the $n^{O(1)}$-variate and $n$-degree polynomials in $\VP$  by depth 3 circuits of size $2^{o(n\log n)}$. 

\end{corollary}

The result of \cite{gk1998} only says that over fixed-size finite fields, not all the $n^{O(1)}$-variate and $n$-degree polynomials in $\VP$ can be computed by $\depththree$ circuits of size $2^{o(n)}$.

Next, we study the depth 4 complexity of $\NW_{n,\epsilon}(\ex)$ polynomial over any field. 
We prove that any depth four $\depthfoursqrtn$ circuit that computes $\NW_{n,\epsilon}(\ex)$ polynomial must be of size $2^{\Omega(\sqrt{n}\log n)}$.  Matching their lower bound, the polynomials considered by Kayal et.al. \cite{kss2013} and Fournier et.al. \cite{flms2013} have depth four $\depthfoursqrtn$ circuits of size $2^{O(\sqrt{n}\log n)}$. In contrast, the polynomial $\NW_{n,\epsilon}(\ex)$ has no known matching upper bound.  
This result is obtained by the application of a key theorem from our recent work \cite{cm2013}. 
 
 \begin{theorem}\label{depth-4-lb}
 For $\epsilon\in (0,1)$, any depth four $\depthfoursqrtn$ circuit computing the $\NW_{n,\epsilon}(\ex)$ polynomial (over any field) must be of size $2^{\Omega_{\epsilon}({\sqrt{n}\log n})}$. 
\end{theorem}

In a very recent work (and independent of ours), Kumar and Saraf have proved super polynomial circuit size lower bound for homogeneous depth 4 circuits (with \emph{no fan-in restriction}) computing the $\NW_{n,\epsilon}(\ex)$ polynomial \cite{ks12013}. 
 
 \subsection*{Proof Idea}
\label{Subsection:ProofIdea}

Our proof technique is quite simple and it borrows ideas mostly from the proof technique of Grigoriev and Karpinski\cite{gk1998}.  $\depththree$ circuits over fixed-size finite fields enjoy a nice property that the derivatives of the high rank product gates can be eliminated except for a few erroneous points (denoted by $E$). This property was first observed by Grigoriev and Karpinski in \cite{gk1998}. To do that they fixed a threshold for the rank of the product gates. Since they were looking for a $2^{\Omega(n)}$ lower bound for the Determinant of 
 a $n\times n$ matrix and the rank of the entire derivative space of of the determinant polynomial is $2^{O(n)}$, 
 it was natural for them to fix the threshold to be $O(n)$. We choose the threshold for the rank of the product gates to be $O(n\log n)$. This allows us to bound the size of the error set meaningfully. 
 
 The dimension of the derivative spaces of the polynomial families $\{\NW_{n,\epsilon}(\ex)\}_{n>0}$ and $\{\IMM_{n,n}(\ex)\}_{n>0}$ is $2^{\Omega(n\log n)}$. We carefully choose subspaces of the derivative spaces of these polynomials that have an additional structure. These subspaces are spanned by a \emph{downward closed} set of  monomials.  Let $\fq$ be the finite field and $N$ be the number of the variables in the polynomial 
 under consideration. The basic idea is to prove that the dimension of the derivative space 
 $H$ of the polynomial being considered is more than the dimension of the set of functions in $H$ which do not evaluate to zero over the entire space $\fq^N$.  Since the subspace $H$ contains only multilinear polynomials, we can then conclude that a nonzero multilinear polynomial in 
 $H$ will evaluate to zero on entire $\fq^N$, which is not possible.  Since we can only bound the rank of the 
 derivative space of $\depththree$ circuits over $\fq^N\setminus E$, we need an argument to lift the upper bound from $\fq^N\setminus E$ to the entire space. We do this by defining a suitable linear map from $H$ to $H$. The 
 downward closed structure of the generator set of $H$ is crucial for the map to be well defined. The argument is motivated by a group symmetry argument given in \cite{gk1998}. In \cite{gk1998}, the argument was on the space of invertible matrices $\GL_n(\fq)$. The fact that the derivative space of the determinant polynomial of a $n\times n$ matrix is invariant under $\GL_n(\fq)$ action was crucially used.   

Finally, we prove Theorem \ref{depth-4-lb} using the method of shifted partial derivatives. We use a 
 key theorem from \cite{cm2013} which was used to unify the lower bound proofs of \cite{kss2013} and 
 \cite{flms2013}.  To apply the result from \cite{cm2013}, we show that for the $\NW_{n,\epsilon}(\ex)$ polynomial, a set of $2^{O(\sqrt{n}\log n)}$ of derivatives of order $O(\sqrt{n})$ have a pair-wise good distance between their leading monomials. 
 
\section{Organization} 
In section \ref{prelim}, we introduce the preliminaries related to arithmetic circuits, partial derivatives, and define the polynomial families $\{\NW_{n,\epsilon}(\ex)\}_{n>0}$ and 
$\{\IMM_{n,n}(\ex)\}_{n>0}$.  
We recall known results related to the derivative space of $\depththree$ circuits in Section~\ref{Section:der-depth3}. In section \ref{Section:Der}, we study the derivative spaces of our polynomial families. We prove  Theorem \ref{main-thm} in Section~\ref{Section:Covering}. In the section \ref{Section:DepthFourBound}, we prove the depth 4 lower bound for the $\NW_{n,\epsilon}(\ex)$ polynomial. We conclude in Section \ref{conclusion}.  
  
\section{Preliminaries}\label{prelim}
\subsubsection*{Arithmetic Circuits}
An arithmetic circuit over a field $\F$ with the set of variables $x_1,x_2,\dots,x_n$ is a directed acyclic graph such that the internal nodes are labelled by addition or multiplication gates and the leaf nodes are labelled by the variables or the field elements. The node with fan-out zero is the output gate. An arithmetic circuit computes a polynomial in the polynomial ring $\F[x_1,x_2,\dots, x_n]$. Size of an arithmetic circuit is the number of nodes and the depth is the length of a longest path from the root to a leaf  node.  

\subsubsection*{Depth 3 Circuits}
Usually a depth 3 circuit over a field $\F$ is denoted by $\depththree$.  The circuit has an addition gate at the top, a middle layer of multiplication gates, and then a level of addition gates at the bottom. A $\Sigma\Pi\Sigma$ circuit with $s$ multiplication gates computes a polynomial of the following form. 
\begin{align*}
  C(\ex)=\sum_{i=1}^s\prod_{j=1}^{d_i}\el_{i,j}(x_1,\dots,x_n)
\end{align*}
where $\el_{i,j}$s are affine linear functions over $\F$ and $\{x_1,x_2,\ldots,x_n\}$ are the variables appearing  in the polynomial.

\subsubsection*{Depth 4 Circuits}
A depth four circuit over a field $\F$ is denoted by $\depthfour$.  It has 
alternating layers of addition and multiplication gates where the top gate is an addition gate. A depth four circuit computes a polynomial of the following form.  
\begin{align*}
C (\ex)= \sum_{i=1}^s\prod_{j=1}^{d_i}Q_{i,j}(x_1,x_2,\dots,x_n)
\end{align*} 
where $Q_{i,j}$s are polynomials in $\F[x_1,x_2,\dots,x_n]$. 
A $\depthfourparam$ circuit is a depth four circuit where the fan-in of the multiplication gates in the bottom layer is bounded by a parameter $t$ and the fan-in of all the multiplication gates in the layer adjacent to the output gate is bounded by the parameter $D$. These circuits compute polynomials of the form $C = \sum_{i=1}^s\prod_{j=1}^{d_i}Q_{i,j}(\ex)$ where the degree of the polynomial $Q_{i,j}$ is bounded by $t$ for all $i$ and $j$, and $d_i\leq D$ for all $i$.  
Building on the results of \cite{av2008} and \cite{koi2012}, Tavenas\cite{tav2013} proved 
the following important theorem. 

\begin{theorem}[Theorem 4, \cite{tav2013}]\label{tav2013}
Let $f$ be an $n$-variate polynomial computed by a circuit of size $s$ and of degree $d$. Then, $f$ is computed by a $\depthfourparam$ circuit $C$ of size $2^{O(\sqrt{d \log(ds) \log n)}}$ where $D=O(\sqrt{d})$ and $t\leq\sqrt{d}$. Furthermore, if $f$ is homogenous, it will also the case for $C$.  
\end{theorem}
 
 \subsubsection*{Partial Derivatives}
 
 For a monomial $\mathbf{x}^{\mathbf{i}} = x_1^{i_1}x_2^{i_2}\dots x_n^{i_n}$, let $\partial^{\mathbf{i}}f$ be the partial derivative of $f$ with respect to the monomial $\mathbf{x}^{\mathbf{i}}$. The degree of the monomial is denoted by $|\mathbf{i}|$ where 
$|\mathbf{i}|:=(i_1 + i_2 + \dots + i_n)$.  We recall the following definition of partial derivative space  from \cite{gkks2012}.
\begin{definition}
  \label{shiftedPartialDerDefn}
  Let $f(\ex)\in \mathbb{F}[\ex]$ be a multivariate polynomial. The span of the $k$-th order derivatives of $f$, denoted by $\langle\partial^{=k}f\rangle$, is defined as
  \begin{align*}
    \langle\partial^{=k}f\rangle= \F\mbox{-span} \{\partial^{\mathbf{j}}f : \mathbf{j} \in \mathbb{Z}^n_{\geq 0} \mbox{ with } |\mathbf{j}| = k \}
  \end{align*}
  We denote by $\dim(\langle\partial^{=k}f\rangle)$ the dimension of the vector space $\langle\partial^{=k}f\rangle$.
\end{definition}

Let $\succ$ be any admissible monomial ordering. 
The \emph{leading monomial} of a polynomial $f(\ex) \in \F[\ex]$, denoted by $\LM(f)$ is the largest monomial $\mathbf{x}^{\mathbf{i}}\in f(\ex)$ under the order $\succ$.

\subsubsection*{Combinatorial Nullstellensatz} 

We recall the following theorem from \cite{Alon99}. 
\begin{theorem}\label{thm:alon}
Let $f(x_1,x_2,\dots,x_n)$ be a polynomial in $n$ variables over an arbitrary field $\F$.  Suppose that the degree of $f$ as a polynomial in $x_i$ is at most $t_i$, for $1\leq i\leq n$ and let $S_i\subseteq \F$ such that $|S_i|\geq t_i + 1$. If $f(a_1,a_2,\dots,a_n)=0$ for all $n$-tuples in $S_1\times S_2\times\dots\times S_n$, then $f=0$.    
\end{theorem}

\subsection{The Polynomial Families}
\label{Section:poly-family} 
A multivariate polynomial family $\{f_n(\ex)\in\F[x_1,x_2,\dots,x_n] : n\geq 1\}$ is in the class $\VP$ if $f_n$ has degree at most $\poly(n)$ and can be computed by an arithmetic circuit of size $\poly(n)$.  
It is in $\VNP$ if it can be expressed as 
\[
f_n(\ex)=\sum_{\ey\in\{0,1\}^m} g_{n+m}(\ex,\ey)
\]
where $m=|\ey|=\poly(n)$ and $g_{n+m}$ is a polynomial in $\VP$.

\subsubsection*{The Polynomial Family from the Combinatorial Design}
\label{Subsection:NW}
Let $\F$ be any field\footnote{In the lower bound proof for $\NW_{n,\epsilon}(\ex)$, we will consider $\F$ to be any fixed finite field $\fq$ such that $q\geq 3$.}. For integers $n>0$ ranging over prime powers and $0<\epsilon<1$, we define a polynomial family $\{\NW_{n,\epsilon}(\ex)\}_{n>0}$ in $\fq[\ex]$ as follows. 
\begin{align*} 
  \NW_{n,\epsilon}(\ex) = \sum_{a(z) \in \F_n[z]}x_{1a(1)} x_{2a(2)}\ldots x_{na(n)} 
\end{align*} 
where $a(z)$ runs over all univariate polynomials of degree $ < \epsilon n$. 
The finite field $\F_n$ is naturally identified with the numbers $\{1,2,\ldots,n\}$. Notice that the number of monomials in $\NW_{n,\epsilon}(\ex)$ is $n^{\epsilon n}$. From the explicitness of the polynomial, it is clear that $\{\NW_{n,\epsilon}(\ex)\}_{n>0}$ is in $\VNP$ for any $\epsilon\in (0,1)$. In \cite{kss2013}, a very similar family of polynomials was considered where the degree of the univariate polynomial was bounded by $\epsilon \sqrt{n}$. 
 

\subsubsection*{The Iterated Matrix Multiplication Polynomial} 
\label{Subsection:IMM}
The iterated matrix multiplication polynomial of $n$ generic $n\times n$ matrices $\ex^{(1)}, \ex^{(2)}, \ldots, \ex^{(n)}$ is the $(1,1)$th entry of the product of the matrices. More formally, let $\ex^{(1)}, \ex^{(2)}, \ldots, \ex^{(n)}$ be $n$ generic $n\times n$ matrices with disjoint sets of variables and $x_{ij}^{(k)}$ be the variable in $\ex^{(k)}$ indexed by $(i,j)\in[n]\times[n]$.
Then the iterated matrix multiplication polynomial (denoted by the family $\{\IMM_{n,n}(\ex)\}_{n>0}$) is defined as follows.

\begin{align*}
  \IMM_{n,n}(\ex) = \sum_{i_1, i_2, \ldots, i_{n-1} \in [n]}x_{1i_1}^{(1)}x_{i_1i_2}^{(2)}\dots x_{i_{(n-2)}i_{(n-1)}}^{(n-1)}x_{i_{(n-1)}1}^{(n)}
\end{align*}
Notice that $\IMM_{n,n}(\ex)$ is a $n^2(n-2) + 2n$-variate polynomial of degree $n$. For our application, we consider $n=2m$ where $m$ ranges over the positive integers. Over any field 
$\F$, the polynomial family $\{\IMM_{n,n}(X)\}_{n>0}$ can be computed in $\VP$. This can be seen by observing that $\IMM_{n,n}(X)$ can be computed by a $\poly(n)$ sized algebraic branching program.

\section{The Derivative Space of $\depththree$ Circuits Over Small Fields}
\label{Section:der-depth3}  


In this section we fix the field $\F$ to be a fixed-size finite field $\fq$ where $q\geq 3$.
Let $C$ be a $\depththree$ circuit of top fan-in $s$ computing a $N=n^{O(1)}$-variate polynomial of 
degree $n$.   Consider a $\Pi$ gate $T=\el_1\el_2\dots\el_d$. Let $r$ be the rank of the (homogeneous)-linear system corresponding to $\{\el_1, \el_2, \dots, \el_d\}$ by viewing  each $\el_i$ as a vector in $\fq^{N + 1}$.  Fix a threshold for the rank of the system of linear functions $r_0 = \beta n\ln n$,  where $\beta > 0$ is a constant to be fixed in the analysis.  In our application, the parameter $N$ is at least $n^2$, so the threshold for the rank is meaningful.  W.l.o.g, let  $\{\el_1, \el_2, \dots, \el_r\}$ be a set of affine linear forms whose homogeneous system forms a maximal independent set of linear functions. The following analysis has been reworked from \cite{gk1998} to fix the parameters. It shows that the derivative space of a $\depththree$ circuit can be approximated by the low rank gates of the circuit over a large subset of  $\fq^N$.
  
 
\subsubsection*{Low rank gates : $r \leq r_0$} 
\label{Subsection:LowRankGates}
Over the finite field $\fq$, we have $x^q = x$. We express $T: \fq^{N}\rightarrow \fq$ as a linear combination of $\{\el_1^{e_1} \el_2^{e_2}\dots \el_r^{e_r} : e_i< q ~ \mbox{for all}~ i\in [r]\}$. Since, the derivatives of all orders lie in the same space, the dimension of the set of partial derivatives of $T$ of all orders is bounded by  $q^{r}\leq q^{r_0}$. 
 
\subsubsection*{High rank gates : $r > r_0$} 
\label{Subsection:HighRankGates}
Let the rank of a high rank gate $T$ be $y\beta n\ln n$ where $y\geq 1$.  
We assign values to the variables uniformly at 
random from $\fq$ and compute the probability that at most $n$ linearly independent functions  evaluate to zero. 
 
\begin{align*} 
  \mbox{Pr$_{a\in\fq^{N}}$[ at most $n$ linearly independent functions evaluate to zero]} &= \sum_{i=0}^n{r \choose i}\left(\frac{1}{q}\right)^i\left(1 - \frac{1}{q}\right)^{r-i}\\ 
  &\leq n{r \choose n}\left(\frac{1}{q}\right)^n\left(1 - \frac{1}{q}\right)^{r-n} 
\end{align*} 

The above inequality follows  from the fact that $r> 2n$. 
Hence, if we differentiate $T$ with respect to any set of variables of size at most $n$ and restrict the variables 
to values from $\fq$, the gate $T$ may not vanish over a set of points $E_T$ whose size is estimated below.  
\begin{align*} 
  |E_T| \leq n{r \choose n}\left(\frac{1}{q}\right)^n\left(1 - \frac{1}{q}\right)^{r-n}q^{N} 
\end{align*} 
 
Over all the gates, let $E$ be the set of points over which some of the product gates with large rank may not evaluate to zero. Then by a union bound, we get that $|E|\leq s |E_T|$. 
If $s\leq e^{\delta n \ln n}$ then we have the following estimate.  
\begin{align*} 
  |E| &\leq e^{\delta n \ln n}n{r \choose n}\left(\frac{1}{q}\right)^n\left(1 - \frac{1}{q}\right)^{r-n}q^{N}\\ 
  &\leq e^{\delta n \ln n}n\left(\frac{er}{n}\right)^ne^{-\frac{r-n}{q}}q^{N}\\ 
  &= q^{N}e^{\delta n \ln n + n+ n\ln\frac{r}{n} + \ln n - \frac{r-n}{q}}\\ 
  &= q^{N}e^{\delta n \ln n + n+ n\ln\frac{y\beta n \ln n}{n} + \ln n  
- \frac{y\beta n\ln n-n}{q}}  
\end{align*} 
To bound the above estimate meaningfully, we need $\delta n \ln n$ to be strictly less than $\frac{y\beta n}{q} \ln n - n\ln y$. That is,  
\begin{align*}
  \delta < \frac{y\beta}{q} - \frac{\ln y}{\ln n} \numberthis \label{eq:ErrorSetEstimate}
\end{align*}
 
 Once we satisfy the relation given by the inequality \ref{eq:ErrorSetEstimate}, we can upper bound the size of $E$ as 
 $|E|<q^{N} \mu^{n\ln n}$ for some suitably fixed constant $\mu$ between $0$ and $1$.  Now it is clear that over $\fq^N \setminus E $, the derivative 
 space is spanned by the derivatives of the low rank gates. We summarize it in the following lemma. 
 
 \begin{lemma}\label{depth3-restriction}
 Let $\fq$ be a fixed-size finite field where $q\geq 3$. Then there exist constants  $0<\delta(q),\beta(q), \mu(q)<1$ such that the following is true.  Let $C$ be a $\depththree$ circuit of top fan-in $s \leq e^{\delta n\ln n}$ computing a $N=n^{O(1)}$-variate and $n$-degree polynomial $f(\ex)$ over the finite field $\fq$. Then, there exists a set $E\subset \fq^N$ of size at most $q^N \mu^{n\ln n}$ such that the dimension of the space spanned by the derivatives of order $\leq n$ of $C$ restricted to $\fq^N\setminus E$ is $\leq s~q^{\beta n\ln n}$.  
 \end{lemma}
 
 In Section \ref{Section:Covering}, we show how to fix the parameters $\delta, \beta$, and $\mu$ which depend only on the field size $q$. 

\section{Derivative Spaces of the Polynomial Families}
\label{Section:Der}

In this section, we study the derivative spaces of $\NW_{n,\epsilon}(\ex)$ and $\IMM_{n,n}(\ex)$ polynomials. Instead of considering the full derivative spaces, we focus on a set of carefully chosen derivatives and consider the subspaces spanned by them.
 
\subsection*{The derivative space of $\{\NW_{n,\epsilon}(\ex)\}_{n>0}$ polynomial family}
\label{Subsection:Der-NW} 
A set of variables $D=\{x_{i_1j_1}, x_{i_2j_2},\dots,x_{i_tj_t}\}$ is called an admissible set if 
$i_k$s (for $1\leq k\leq t$) are all distinct and $\epsilon n \leq t \leq n$. 
Let $H$ be the subspace spanned  by the set of the partial derivatives of the polynomial $\NW_{n,\epsilon}(\ex)$ with respect to the admissible sets of variables. More formally, 
 
\begin{align*} 
  H := \fq\mbox{-span}\left\{\frac{\partial \NW_{n,\epsilon}(\ex)}{\partial D} : D~\mbox{is an admissible set of variables}\right\} 
\end{align*} 
 
Since the monomials of the $\NW_{n,\epsilon}(\ex)$ polynomial are defined by the univariate polynomials of degree $< \epsilon n$, each partial derivative with respect to such a set $D$ yields a multilinear monomial. If we choose $\epsilon$ such that $n-\epsilon n > \epsilon n$ (i.e. $\epsilon < 1/2$), then  after  the differentiation, all the monomials of length $n-\epsilon n$ are distinct. This follows from the fact that  the monomials are generated from the image of the univariate polynomials of degree $<\epsilon n$. 

Let us treat these monomials as functions from $\fq^{n^2}\rightarrow\fq$. The following lemma says that the functions corresponding to any set of distinct monomials are linearly independent. 

\begin{lemma}\label{comb-null}
Let $m_1(\ex), m_2(\ex), \dots, m_k(\ex)$ be any set of $k$ distinct monomials in
$\F_q [x_1,x_2,\dots,x_N]$. For $1\leq i\leq k$, let $f_i : \F_q^N\rightarrow \F_q$ be the function corresponding to the monomial $m_i(\ex)$, i.e. $f_i(\ex)=m_i(\ex)$. Then, $f_i$s are linearly independent in the  $q^N$ dimensional vector space over $\F_q$. 
\end{lemma}

\begin{proof}
If $f_i$s are not linearly independent then $\sum_{i=1}^k \lambda_i f_i = 0$ for 
$\bar{\lambda}=(\lambda_1, \lambda_2,\dots,\lambda_k)\in \F_q^k\setminus\{\bar{0}\}$. Then, the nonzero 
multilinear polynomial $\sum_{i=1}^k \lambda_i m_i(\ex)$ evaluates to zero on $\F_q^N$, which contradicts Theorem \ref{thm:alon}.  
\end{proof}

Consider the derivatives of $\NW_{n,\epsilon}(\ex)$ corresponding to the sets 
$\{x_{1 a(1)}, x_{2 a(2)}, \dots, x_{\epsilon n a(\epsilon n)}\}$ for all univariate polynomials $a$ of degree $<\epsilon n$. 
From Lemma \ref{comb-null}, it follows that  $\dim(H) \geq n^{\epsilon n}=e^{\epsilon n \ln n}$.  W.l.o.g, we can assume that the constant function $\mathbf{1} : \fq^{n^2}\rightarrow \fq$ given by 
$\forall x, \mathbf{1}(x)=1$ is also in $H$. This corresponds to the derivatives of order $n$.  

\subsection*{The derivative space of $\{\IMM_{n,n}(\ex)\}_{n>0}$ polynomial family}
\label{Subsection:Der-IMM}
For our application, we consider $n=2m$ where $m$ ranges over the positive integers. Consider the set of matrices $\ex^{(1)},  \ex^{(3)}, \ldots, \ex^{(2m-1)}$ corresponding to the odd places. Let $S$ be any set of $m$ variables chosen as follows. Choose any variable from the first row of $\ex^{(1)}$ and choose any one variable from each of the matrices $\ex^{(3)}, \ldots, \ex^{(2m-1)}$.  We call such a set $S$  an admissible set.  

If we differentiate $\IMM_{n,n}(\ex)$ with respect to two different admissible sets of  variables  $S$ and $S'$,  then we get  two different monomials of length $m$ each. This follows  from the structure of the monomials in the $\IMM_{n,n}(\ex)$ polynomial, whenever we fix two variables from $\ex^{(i-1)}$ and $\ex^{(i+1)}$, the variable from $\ex^{(i)}$ gets fixed. So the number of such monomials after differentiation is exactly $n^{2m-1}= e^{(n-1)\ln n}$. 

Let $m_{S}$ be the monomial obtained after differentiating $\IMM_{n,n}(\ex)$ by the set of variables in $S$ and $\var(m_S)$ be the set of variables in $m_S$. Consider the derivatives of  $\IMM_{n,n}(\ex)$ with respect to the following sets of variables.

\begin{align*}
  \{ S \cup T : T \subseteq \var(m_S) \}~ \mbox{where $S$ ranges over all admissible sets}. 
\end{align*}

Let $H$ be the subspace spanned by these derivatives. More formally, 

\begin{align*} 
  H := \fq\mbox{-span}\left\{\frac{\partial \IMM_{n,n}(X)}{\partial D} : D =S \cup T 
  ~\mbox{where}~T\subseteq \var(m_S)  ; S ~\mbox{is an admissible set}\right\}  
\end{align*}

As before, we can assume that the constant function $\mathbf{1}$ is in $H$.  
From Lemma \ref{comb-null},  we know that $\dim(H)\geq e^{(n-1)\ln n}$. 
Now to unify the arguments for $\NW_{n,\epsilon}(\ex)$ and $\IMM_{n,n}(\ex)$ polynomials, we introduce the following notion.

\subsubsection*{Downward closed property} 
\begin{definition}
 A set of monomials $\M$ is said to be downward closed if the following property holds. If $m(\ex) \in \M$ and $m'(\ex)$ is such that $\var(m'(\ex)) \subseteq \var(m(\ex))$, then $m'(\ex) \in \M$.
\end{definition}
 
 Now we consider a downward closed set of monomials $\M$ over $N$ variables. These monomials can be viewed as functions from $\fq^N$ to $\fq$. W.l.o.g, we assume that the constant function is also in $\M$ (constant function corresponds to a monomial with an empty set of variables). Let $H$ be the subspace spanned by these functions in $\M$. 

For any $u\in \fq^{N}$, define an operator $T_u$ such that $(T_u(f))(\ex) = f(\ex-u)$ for any function $f:\fq^{N}\rightarrow\fq$. The following proposition is simple to prove.  
 
\begin{prop} 
Let $H$ be the subspace spanned by a downward closed set of monomials $\M$ over the set of variables 
$\{x_1, x_2, \ldots, x_N\}$. Then for any $u\in \fq^N$, $T_u$ is a linear map from $H$ to $H$. 
\end{prop} 
\begin{proof} 
  Let $g(\ex)$ be an arbitrary function in $H$ which can be expressed as follows: $g(\ex) =\sum_{i\geq 1}c_im_i(\ex)$  where $m_i(\ex) \in \M$, and $c_i \in \fq$ for all $i\geq 1$.
  \begin{align*}
    (T_u(g))(\ex) &= g(\ex-u) = \sum_{i\geq 1}c_im_i(\ex-u) 
  \end{align*}

  It is sufficient to prove that $m(\ex-u)\in H$ where $m(\ex) \in \M$. We can express $m(\ex-u)$ as follows.
  \begin{align*}
    m(\ex-u) = \sum_{S\subseteq~\var\left(m\left(\ex\right)\right)} c_S\prod_{x_r\in S}x_r
  \end{align*}
  where $c_S\in \fq$. For every $S\subseteq~\var(m(\ex))$,  ${\prod_{x_r\in S}} x_r\in \M$ because $\M$ is downward closed. Since the choice of $S$ was arbitrary, $m(\ex-u)\in H$.  It is obvious that $T_u$ is a linear map. 
\end{proof} 
 
It is not difficult to observe that the derivative spaces that we select for $\NW_{n,\epsilon}(\ex)$ and $\IMM_{n,n}(\ex)$ are spanned by  downward closed sets of monomials.

\begin{lemma}\label{down-closed}
The generator sets for the derivative subspaces $H$ for $\NW_{n,\epsilon}(\ex)$ and $\IMM_{n,n}(\ex)$ polynomials are downward closed. 
\end{lemma}

\begin{proof}
Let us consider the $\NW_{n,\epsilon}(\ex)$ polynomial first. Let $m \in H$ be any monomial and $D$ 
be the admissible set such that $m=\frac{\partial \NW_{n,\epsilon}(\ex)}{\partial D}$. Let $m'$ be any monomial such that $\var(m')\subseteq \var(m)$.  Then $m'=\frac{\partial \NW_{n,\epsilon}(\ex)}{\partial D'}$ where $D'=D\cup (\var(m)\setminus \var(m'))$. 

Similarly for the $\IMM_{n,n}(\ex)$ polynomial, consider any $m\in H$.  Then $m=\frac{\partial \IMM_{n,n}(X)}{\partial D}$ and $D=S\cup T$ for an admissible set $S$ and $T\subseteq\var(m_S)$.  
If $m'$ is any monomial such that $\var(m')\subseteq \var(m)$, then $m'=\frac{\partial \IMM_{n,n}(X)}{\partial D'}$ where $D'=S\cup (T\cup (\var(m)\setminus\var(m')))$.  Clearly $T\cup 
(\var(m)\setminus\var(m'))\subseteq \var(m_S)$. 
\end{proof}

\section{A Covering Argument} 
\label{Section:Covering}
In this section, we adapt the covering argument of  \cite{gk1998} to prove the lower bound results. 
In \cite{gk1998}, the covering argument was given over 
the set of invertible matrices. Here we adapt their argument suitably over the entire space $\fq^N$.  As defined in the section \ref{Section:Der}, the subspace $H$ represents the chosen derivative subspace of either the $\NW_{n,\epsilon}(\ex)$ polynomial or the $\IMM_{n,n}(\ex)$ polynomial. 

Define the subspace $H_a := \{f\in H : f(a)=0\}$ for $a \in \fq^{N}$. Let us recall that $E$ is the set of points over which some of the product gates with large rank may not evaluate to zero. Let the set of points $\fq^N\setminus E$ be denoted by $A$. Then $\bigcap_{a\in A}H_a$ denotes the set of functions which evaluate to zero over all points in $A$. Now, we consider the space of the functions which do not evaluate to zero over all of $A$. From Lemma  \ref{depth3-restriction}, we get that $\codim(\bigcap_{a\in A}H_a) < s~q^{r_0}$.  

\begin{prop} 
  For any $u , a \in \fq^{N}$, we have that $T_u(H_a) = H_{u+a}$. 
\end{prop} 
\begin{proof} 
Let $f(\ex)$ be any function in $H_a$.  $(T_u(f))(\ex) = f(\ex-u)$. Since $f(a) = 0$, $f(a+u-u)=0$. Also the monomials of $f(\ex-u)$ are obtained as the subsets of the monomials of $f$. Since $H$ is generated by a downward closed set of monomials, it is clear that $f(\ex-u)\in H$ implying  $f(\ex-u)\in H_{u+a}$.  So $T_u(H_a)\subseteq H_{u+a}$. 

Now consider any function $g(\ex)\in H_{u+a}$. Define the function $h(\ex)=g(\ex + u)$. Then 
$h(a)=0$ and so $h(\ex)\in H_a$ \footnote{The fact that $h(\ex)\in H$ follows again from the downward closed property of the generators.}.  Also, $T_u(h(\ex)) = g(\ex)$. Hence, $H_{a+u}\subseteq T_u(H_a)$.    
\end{proof} 
 
Let $P = \bigcap_{a\in A}H_a$. Let $S\subset \fq^{N}$ be a set such that we can 
cover the entire space $\fq^N$ by the shifts of $A$ with the elements from $S$.

\begin{align*}
\bigcup_{u\in S}u + A = \fq^{N} 
\end{align*} 
 Now by applying the map $T_u$ to $P$, we get the following. 
 \begin{align*}
  T_u(P) = \bigcap_{a\in A}T_u(H_a) = \bigcap_{b\in u+A}H_b
  \end{align*} 
    
    By a further intersection over $S$, we get the following. 

\[
  \bigcap_{u\in S} T_u(P) = \bigcap_{u\in S}\bigcap_{b\in u+A}H_b = \bigcap_{b\in\fq^{N}}H_b \numberthis \label{coveringstatement} 
\] 

From Equation~\ref{coveringstatement}, we get the following estimate.
\begin{align*} 
  \codim\left(\bigcap_{b\in\fq^{N}}H_b\right) = \codim\left(\bigcap_{u\in S}T_u(P)\right) \leq |S|~\codim(P) \leq |S|~ s~ q^{r_0} \numberthis\label{eq:codim-estimate}
\end{align*} 
The $\codim\left(\bigcap_{b\in\fq^{N}}H_b\right)$ refers to the dimension of the set of functions in $H$ which do not evaluate to zero over all the points in $\fq^{N}$. 
 
Next, we show an upper-bound estimate for the size of the set $S$.  This follows from a simple 
adaptation of the dominating set based argument given in \cite{gk1998}. 

\subsection*{Upper bound on the size of the set $S$}
 
Consider the directed graph $G = (V,R)$ defined as follows. The points in $\fq^{N}$ are the vertices of the graph. For $u_1, u_2\in \fq^{N}$, the edge $u_1\rightarrow u_2$ is in $R$ iff $u_2 = u_1+b$ for any $b\in A$. Clearly the in-degree and out-degree of any vertex are equal to $|A|$. 
Now, we recall  Lemma~2 of \cite{gk1998} to estimate the size of $S$. 

\begin{lemma}[\cite{lov1975}]\label{lov75}
Let $(V,R)$ be a directed (regular) graph with $|V| =m$ vertices and with the in-degree and the out-degree of each vertex both equal to $d$. Then there exists a subset $U\subset V$ of a size $O(\frac{m}{d}\log(d+1))$ such that for any vertex $v\in V$ there is a vertex $u\in U$ forming an edge $(u,v)\in R$.
\end{lemma}  
Let $c_0$ be the constant fixed by the lemma in its $O()$ notation.  By Lemma \ref{lov75}, we get the following estimate. 

\begin{align*} 
  |S| &\leq c_0\frac{|\fq^{N}|}{|A|}\log (|A|+1)\\ 
  &\leq c_0 \frac{q^{N}}{q^{N}-|E|}\log (q^{N}-|E|+1)\\ 
  &\leq c_0\log q ~N~\frac{q^{N}}{q^{N}-|E|}\\ 
  &= O(N) 
\end{align*} 
 
 The last equation follows from the estimate for  $|E|$ from the section \ref{Section:der-depth3}.   
\subsubsection*{Fixing the parameters} 
Consider the inequality~\ref{eq:ErrorSetEstimate} which is $\delta < \frac{y\beta}{q} - \frac{\ln y}{\ln n}$. Fix the values for $\beta, \delta$, and $\mu$ in Lemma \ref{depth3-restriction} as follows.  Set $\beta=\frac{1}{10\ln q}, 
\delta = \frac{1}{20 q\ln q}, \nu=\frac{\delta}{2}$, and $\mu=e^{-\nu}$. 
Consider the function $g(y) = y - \frac{10q\ln q}{\ln n}\ln y -0.50$. Since $g(y)$ is a monotonically increasing function which takes the value of $0.50$ at $y=1$, $g(y)>0$ and thus $\delta < \frac{y\beta}{q} - \frac{\ln y}{\ln n}$ for the chosen values of $\beta$ and $\delta$.  
Also, $\frac{y\beta}{q} - \frac{\ln y}{\ln n} - \delta > \nu$ and thus $|E|\leq q^{N}\mu^{n\ln n}$. 

From Section \ref{Section:Der}, we know that $\dim(H)$ for $\NW_{n,\epsilon}(\ex)$ is at least $e^{\epsilon n\ln n}$. Consider the upper bound on $\codim\left(\bigcap_{b\in\fq^{N}}H_b\right)$ given by the inequality \ref{eq:codim-estimate}. If we choose $\epsilon$ in such a way that $e^{\epsilon n\ln n} > |S| ~s~q^{r_0}$, then there 
will be a multilinear polynomial $f$ in $H$ such that $f$ will evaluate to zero over all points in $\fq^N$. 

\begin{align*} 
  \dim(H) &> n^{\epsilon n} = e^{\epsilon n \ln n}\\ 
  \implies e^{\epsilon n\ln n} &> |S| ~ s~ q^{r_0} = e^{\delta n \ln n +(\beta\ln q) n \ln n + \ln N}\\ 
\end{align*}

Considering the terms of the order of $n\ln n$ in the exponent, it is enough to choose $\epsilon (< 1/2)$ such that the following holds. 
\begin{align*}
\epsilon &> \delta + \beta \ln q\\ 
  &= \frac{1}{20 q\ln q} + \frac{1}{10} 
\end{align*} 

Since the $\dim(H)$ for $\IMM_{n,n}(\ex)$ is $\geq e^{(n-1)\ln n}$, the chosen values of $\beta$ and $\delta$ clearly suffice. Finally,  recall from Theorem \ref{thm:alon} that no non-zero multilinear polynomial can be zero over $\fq^N$.
Thus, we get the main theorem (restated from Section \ref{Section:Intro}).
\begin{theorem}
For any fixed-size finite field $\fq$ such that $q\geq 3$, any depth three $\depththree$ circuit computing the polynomials $\NW_{n,\epsilon}(\ex)$ or $\IMM_{n,n}(\ex)$ must be of size at least $2^{\delta n\log n}$ where $\delta,\epsilon\in(0,1)$ and depend only on $q$.  
\end{theorem}

It is straightforward to observe that the lower bound analysis holds for any polynomial 
for which we can find a subspace (of sufficiently large dimension) of its derivative space spanned by a downward closed set of monomials.

\section{Depth 4 Circuit Size Lower Bound for $\NW_{n,\epsilon}(\ex)$ Polynomial} 
\label{Section:DepthFourBound}

In this section we prove the depth 4 size lower bound for the $\NW_{n,\epsilon}(\ex)$ polynomial. This result holds over any field. It was shown in \cite{cm2013} that any polynomial that satisfies a simple combinatorial property called \emph{Leading Monomial Distance Property} would require $2^{\Omega(\sqrt{n}\log n)}$-sized depth four $\depthfoursqrtn$ arithmetic circuits computing it.  
To define the Leading Monomial Distance Property, we first define the notion of distance between two monomials.
\begin{definition}
  \label{def:MonomialDistance}
  Let $m_1, m_2$ be two monomials over a set of variables. Let $S_1$ and $S_2$ be the (multi)-sets of variables corresponding to the monomials $m_1$ and $m_2$ respectively. The distance $\Delta(m_1, m_2)$ between the monomials $m_1$ and $m_2$ is the $\min\{|S_1|-|S_1\cap S_2|, |S_2|-|S_1\cap S_2|\}$ where the cardinalities are the order of the (multi)-sets. 
\end{definition}

For example, let $m_1 = x_1^2x_2x_3^2x_4$ and $m_2 = x_1x_2^2x_3x_5x_6$. Then $S_1 = \{x_1, x_1, x_2, x_3, x_3, x_4\}$, $S_2 = \{x_1, x_2, x_2, x_3, x_5, x_6\}$, $|S_1|=6$, $|S_2|=6$ and 
$\Delta(m_1, m_2) = 3$. 

We say that a $n^{O(1)}$-variate and $n$-degree polynomial has the Leading Monomial Distance Property, if the leading monomials of a \emph{large subset} ($\approx n^{\sqrt{n}}$) of its span of the derivatives (of order $\approx \sqrt{n}$) have \emph{good pair-wise distance}. 
We denote the leading monomial of a polynomial $f(\ex)$ by $\LM(f)$. Let $\langle\partial^{=k}(f)\rangle$ be the subspace spanned by the $k$th order derivatives of $f$ as defined in Section 
\ref{Section:Intro}. 

\begin{theorem}[\cite{cm2013}]
\label{thm:DepthFourAbstract}
  Let $f(\ex)$ be a $n^{O(1)}$-variate polynomial of degree $n$. Let there be at least $n^{\delta k}$ ($\delta$ is any constant $>0$) different polynomials in $\langle\partial^{=k}(f)\rangle$ for $k=\mu \sqrt{n}$ such that any two of their leading monomials have a distance of at least $\Delta\geq \frac{n}{c}$ for any constant $c>1$, and $0<\mu<\frac{1}{40c}$. Then any depth four $\depthfoursqrtn$ circuit that computes $f(\ex)$ must be of size $e^{\Omega_{\delta,c}(\sqrt{n}\ln n)}$. 

\end{theorem}  

We use the above theorem to prove the depth 4 lower bound for the $\NW_{n,\epsilon}(\ex)$ polynomial (Theorem \ref{depth-4-lb}). 
Let us consider the polynomial $\NW_{n,\epsilon}(\ex)$ over any field $\F[\ex]$. First, we fix an ordering on the variables: $x_{11}\succ x_{12} \succ \dots \succ x_{nn}$. We fix a threshold $k=\mu\sqrt{n}$ where $\mu$ is  a constant to be fixed later. 

Consider the sets of variables of the form $\{x_{1 i_1}, x_{2 i_2}, \ldots, x_{k i_k}\}$ such that $i_j\in [n]$ for $1\leq j\leq k$. Corresponding to any such set of variables, we can find a univariate polynomial $a(z)\in\F_n[z]$ of degree $< \epsilon n$ such that $a(j)=i_j$ for $1\leq j\leq k$. Number of such different sets of size $k$ is $n^k$. Let us represent these sets by $S_1, S_2, \dots, S_{n^k}$.

Now, we partition the monomials of  $\NW_{n,\epsilon}(\ex)$  into sets $B_1, B_2, \ldots, B_{n^k}$ such that for $1\leq i\leq k$, each monomial in the set $B_i$ contains all the variables of the set $S_i$.  

Let $P_i(\ex)$ be the polynomial corresponding to the sum of monomials in $B_i$ for all $i\in[n^k]$. We can express the polynomial $\NW_{n,\epsilon}(\ex)$ as follows. 

\begin{align*}
\NW_{n,\epsilon}(\ex)=\sum_{i=1}^{n^k} P_i(\ex)
\end{align*}

Now, if we differentiate the polynomial $\NW_{n,\epsilon}(\ex)$ with respect to any particular set of variables $S_j = \{x_{1 i_1}, x_{2 i_2}, \ldots, x_{k i_k} \}$, we can see that only one of the polynomials in $\{P_1(\ex), P_2(\ex), \ldots, P_{n^k}(\ex)\}$ contributes a leading monomial. Since any two monomials can intersect at at most $\epsilon n - 1$ places, the distance between the leading monomials (that we get after differentiation) will be $\Delta\geq n-\mu\sqrt{n}-\epsilon n$. We can consider any $\epsilon \in (0,1)$, and then fix $c$ to be any constant $\geq \lceil\frac{2}{1-\epsilon}\rceil$. Finally, we fix $\mu$ to any value such that $0<\mu<\frac{1}{40c}$.  

Thus, we get at least $n^k$ different polynomials in $\langle\partial^{=k}(f)\rangle$ such that any two 
of their leading monomials have pair-wise distance $\geq n/c$. 
Now we apply Theorem~\ref{thm:DepthFourAbstract} with $\delta=1$ to obtain the following theorem (Theorem \ref{depth-4-lb}, restated from Section \ref{Section:Intro}). 

\begin{theorem}
Any depth four $\depthfoursqrtn$ circuit for computing the $\NW_{n,\epsilon}(\ex)$ polynomial (over any field) must be of size $2^{\Omega_{\epsilon}({\sqrt{n}\log n})}$. 
\end{theorem}

In \cite{kss2013} and \cite{flms2013}, the polynomials for which the similar lower bounds were proved have matching upper bounds of $2^{O(\sqrt{n}\log n)}$.  From \cite{gkks2012}, the current depth 4 size lower bounds for  Determinant and Permanent are  $2^{\Omega(\sqrt{n})}$.  To the best of our 
knowledge, the polynomial $\NW_{n,\epsilon}(\ex)$ is the first example of an explicit polynomial in $\VNP$ for which we have the depth four $\depthfoursqrtn$ circuit size lower bound of $2^{\Omega(\sqrt{n}\log n)}$ but no known matching upper bound.  

\section{Conclusion}\label{conclusion}

Then main interesting open problem is to prove that over the fixed-size fields, 
any $\depththree$ circuit computing the determinant polynomial for  a
$n\times n$ matrix must be of size $2^{\Omega(n\log n)}$. For an optimist, 
the task will be to find a $\depththree$ circuit of size $2^{o(n\log n)}$ 
for the determinant polynomial. It seems that we need significantly new 
ideas and techniques to make progress either on the lower bound side or on the 
upper bound side.

\bibliographystyle{alpha} 
\bibliography{ref1} 
 
\end{document}